# SECP-Net: SE-Connection Pyramid Network of Organ At Risk Segmentation for Nasopharyngeal Carcinoma

**Zexi Huang (1), Lihua Guo (1), Xin Yang (2), Sijuan Huang (2)**
((1) School of Electronic and Information Engineering, South China University of Technology,
(2) Sun Yat-sen University Cancer Center)

## Abstract

Nasopharyngeal carcinoma (NPC) is a kind of malignant tumor. Accurate and automatic segmentation of organs at risk (OAR) of computed tomography (CT) images is clinically significant. In recent years, deep learning models represented by U-Net have been widely applied in medical image segmentation tasks, which can help doctors with reduction of workload and get accurate results more quickly. In OAR segmentation of NPC, the sizes of OAR are variable, especially, some of them are small. Traditional deep neural networks underperform during segmentation due to the lack use of global and multi-size information. This paper proposes a new SE-Connection Pyramid Network (SECP-Net). SECP-Net extracts global and multi-size information flow with se connection (SEC) modules and a pyramid structure of network for improving the segmentation performance, especially that of small organs. SECP-Net also designs an auto-context cascaded network to further improve the segmentation performance. Comparative experiments are conducted between SECP-Net and other recently methods on a dataset with CT images of head and neck. Five-fold cross validation is used to evaluate the performance based on two metrics, i.e., Dice and Jaccard similarity. Experimental results show that SECP-Net can achieve SOTA performance in this challenging task.

**Index Terms--** Segmentation of OAR of Nasopharyngeal Carcinoma, deep learning, se-connection pyramid network, auto-context cascaded network.

## Introduction

Nasopharyngeal Carcinoma (NPC) is a kind of malignant tumor of high incidence in China [1] and its attack rate achieves the first of malignant tumor of ear, nose and throat. If radiotherapy (RT) areas aren't controlled suitably when the patients with this sort tumor are treated by RT, it is possible that normal organs will be affected which has a negative effect on patients' health. Computed Tomography (CT) images are standard resource in RT utilized by doctors for manually segmenting OAR, which makes RT areas are limited in target region strictly. Thus RT won't do damage to normal organs. It shows that segmentation of CT images plays an important role in clinical diagnosis and treatment. However, the workload of manual segmentation is so heavy and time-consuming. If an automatic segmentation method of OAR can be designed in CT images, it will not only reduce doctors' workload, but also be capable of outputting segmentation results in time to save time for patients' treatment.

In recent years, deep learning methods have been widely applied in medical image segmentation. Various models based on convolutional neural network (CNN) have been proposed for medical segmentation tasks and achieved great success. Since proposed, fully convolutional network (FCN) [2] has become basic frame of semantic segmentation. The encoder-decoder structure of FCN is capable of extracting image features from local information to higher spatial features by convolution and pooling measures. Besides, FCN takes advantage of upsampling to get an end-to-end network, and utilizes skip-connection to introduce original feature details fetched by encoder, which improves accuracy of segmentation network. Based on FCN, U-Net [3] proposed in 2015 has some improvement on skip-connection, which fuses features from encoder with corresponding part from decoder in channel dimension to get more refined segmentation results. Thus U-Net has been the most representative network of medical segmentation tasks in recent years. On account of key thoughts of U-Net, many new designs of networks have been presented: Res-U-Net inspired by residual connection [4] replaces submodules in U-Net with residual modules to learn more different features; Dense-U-Net [5] leverage the dense connection to maximally retain information and gradient flow.

Although U-shape networks have achieved remarkable success in the medical image segmentation, they have three aspects of disadvantages.

1) First, the skip-connection in U-Net is too simple because it directly introduces features captured by encoder to decoder instead of non-linear transformation, therefore, it will weaken its learning ability and result in classification errors due to noise.
2) Second, there is no sufficient multi-size information extracted or utilized by U-Net, which does not achieve good performance for objects of complex structure.
3) Third, U-Net lacks use of the global context information and these types of information will be diluted when transmitted to shallower layers.

In the past few years, many new models aiming at improving shortcomings of U-Net have been proposed: Attention U-Net [6] adds a novel self-attention module to skip-connection, and adopts non-linear transformation to enhance learning ability; CE-Net [7] employs various blocks with different receptive field to improve ability of multi-size information extraction; U-Net++ [8] comes up with a kind of pyramid-like network to integrate the information from diverse levels, which makes good use of global and multi-scale context information; CPF-Net [9] not only designs a global pyramid guidance (GPG) module to combine multi-stage global context information, but also imports some convolution blocks with various dilation rate to capture information in different sizes and shapes. Based on our knowledge, there is no any networks which can overcome all three disadvantages in U-Net.

Back to our certain segmentation task, there exists multi-size OAR of NPC. Compared to big organs, small organs' profile isn't clear enough due to its size, thus it's more difficult for segmentation by deep learning networks. In this paper, eight organs will be segment, which are eyes, temporal lobe, mandible, brain stem, parotid, submandibular, thyroid gland and spinal cord. It's obvious that these organs are in different shapes, and there is no doubt that we need a strong network being able to capture multi-size features. As previously mentioned, CPF-Net captures different sizes of features by several dilated convolution kernels of various dilation rate. In origin paper the author chooses at most three different kernels. However, in the segmentation task of OAR of NPC, there are eight organs of different shapes to be segmented. CPF-Net has insufficient parallel dilated convolutions to capture multi-size information from all the organs.

In a word, every of Attention U-Net, CE-Net and U-Net++ only improves one or two disadvantages of U-Net. Though CPF-Net overcomes all the disadvantages, it's not a suitable method for our segmentation task of OAR of NPC due to its intrinsic property.

Inspired by the discussion above, a novel skip-connection module and a pyramid structure are proposed in this paper to overcome the disadvantages mentioned previously. We term the proposed network SE-Connection Pyramid Network (SECP-Net). For the first disadvantage, we design a SE-Connection (SEC) module used for skip-connection. SEC module makes use of the SE block [10] for its non-linearity and channel-wise attention to enhance learnability reduce useless noise influence from input images, respectively. For the second and third disadvantages, we further propose a pyramid network cooperating with SEC module to merge information from multiple stages and get the global context information. In this pyramid structure, the information from each stage consists of local context information from low-level stages and aggerate context information from high-level stages. The network captures features from small size to large size with the progress of the stages from shallow to deep. Capitalize on that, we can directly utilize features of multi-size extracted by the encoder of the network instead of many atrous convolution kernels which are used in CPF-Net. Furthermore, a cascaded network architecture is designed. In this architecture, a deigned pyramid U-Net model is introduced to get the region of interesting, and the probability distribution output from the pyramid U-Net is combined with original input images. Then the combination is sent to an original U-Net regarded as a secondary network to get accurate segmentation images. The novel network is applied for a challenging segmentation task of OAR of NPC.

Our main contributions are summarized in three aspects as follows:

(1) A SE-Connection module is introduced which weakens noise in skip-connection and enhances the non-linearity and learning ability of network. We design pyramid U-Net structure to capture global context information and multi-size information, and fuse them with SEC module to significantly improve segmentation accuracy of small organs. We further present a cascaded network according to auto-context to make the network more focus on ROI.

(2) The proposed SE-Connection module based on U-Net can be easily added into other networks and applied for segmentation task of medical images.

(3) Our method has achieved best performance on the dataset with CT images of head and neck compared to other networks.

The overall structure of this paper takes the form of four parts, including this introductory chapter. In the second part we'll introduce the proposed method in detail. The third part will show the experiment results and discussion. Finally, we get the result in the fourth part.

## Methods

A. Overview

Figure 1 shows the overall structure of SECP-Net, which is a cascaded network. The left part is the primary part of SECP-Net and the right part is the secondary network. The primary network based on a U-shape network consists of three parts: feature encoder, SEC module and decoder.

SEC module is located at the skip-connection between feature encoder and decoder. Coordinated with the pyramid structure, SEC module is able to extract and fuse multi-size information and global context information. SEC module also utilizes attention mechanism to highlight contributing features for segmentation. The secondary network is an original U-Net connected with the left part by auto-context method, which increase the depth of the network to get segmentation more accurate.

B. Encoder and Decoder

Encoder is used for capturing the feature information from input CT images while decoder is used to restore images. In SECP-Net, there are two pairs of encoders and decoders in primary network and secondary network respectively. The structure of them is the same with an original U-Net.

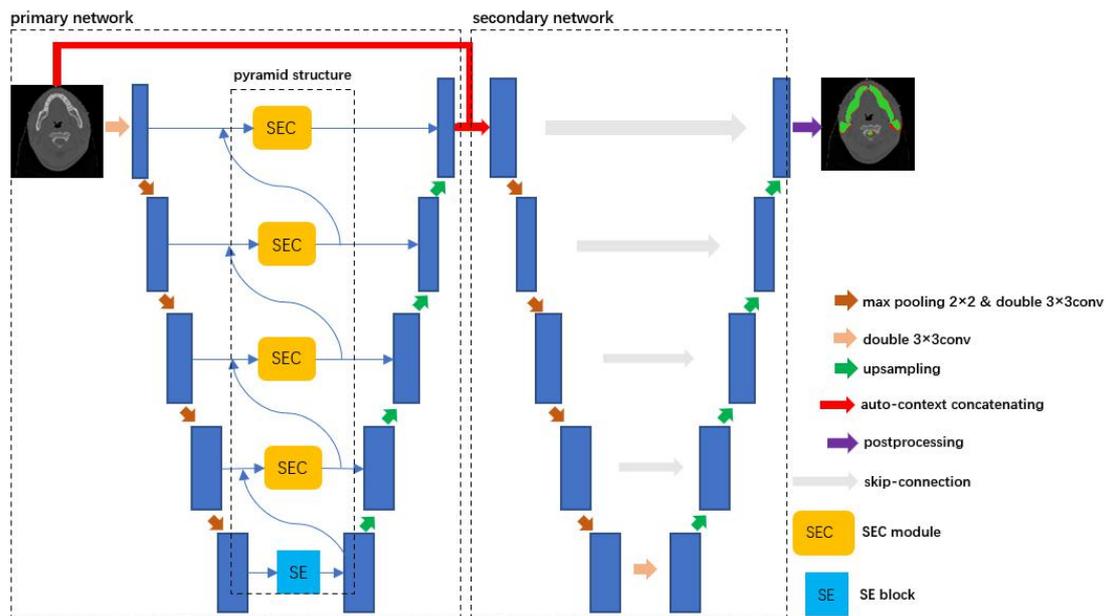

Fig. 1. Overall structure of SECP-Net. After multiple convolution and downsampling, there are different size and level of information captured in different stages for input images. We utilize SEC module and pyramidal network to fuse the information and to extract more contributing features to segmentation by channel attention mechanism of SE block [10]. Then the information flow transmitted to decoder through skip connection. After the extraction of the primary part, we combine original inputs and the probability distribution from the primary network according to auto-context, which is sent to the secondary U-Net for further segmentation and more accurate results.

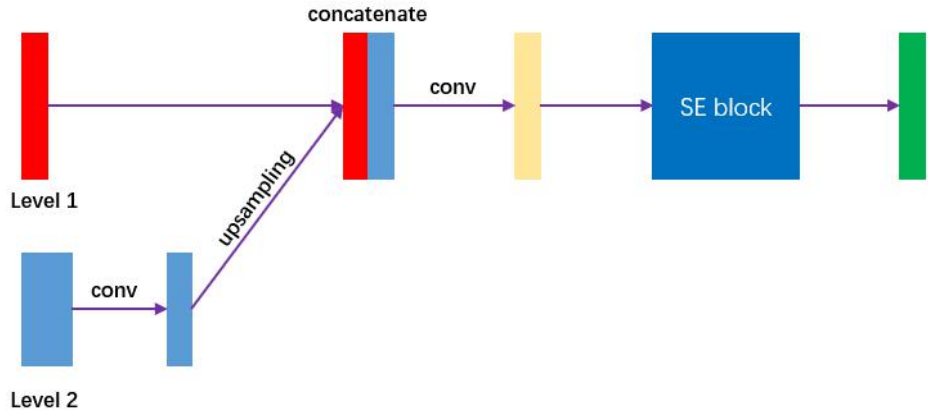

Fig. 2. The illustration of SEC module in a single stage. Multi-size information flow gets fused stage by stage. The fusion information will be sent to SE block for acquisition of more contributing features. Then the output of SE block is sent to decoder by skip connection.

C. SEC module and the Pyramid Framework

For U-shape network, its encoder is able to learn global context information by gradually increasing respective field of convolution kernels, which contains background of segmentation targets and their own features. However, the information flow may be weakened when transmitted to shallower stages in steps of upsampling. U-Net utilizes skip-connection to overcome this disadvantage. But as mentioned previously in the first disadvantage of U-Net, the skip-connection of U-Net is so simple that the learning ability will be weakened due to non-linearity, and this type of structure will introduce noise resulting in ambiguous error for pixel classification. U-Net isn't capable of extracting multi-size information, which makes the model underperform in tasks of segmenting multiple targets according to the second and third disadvantages. In this paper, we present a SEC module and a pyramid network to address above disadvantages. The structure of SEC module is shown in figure 2.

In SEC module, the information between adjacent level stages is fused. As is shown in figure2, Level 1 and Level 2 are feature maps captured in two continuous levels, in which Level 1 is from a shallower stage and Level 2 is the output of SEC module from a deeper stage. The Level 2 is convoluted firstly to keep its channel space the same as Level 1. Next the generated feature maps are upsampled to the same size as Level 1. Then we concatenate Level 1 and processed Level 2 feature maps in channel-wise, and utilize convolution to get fusion results which have half channels of the previous concatenation. Finally, the fusion results are sent to a SE block [10] for purpose of capturing more contributing multi-size features with channel-wise attention mechanism. It's evident that by embedding SEC module between encoders and decoders, the skip-connection has got more non-linearity due to the more complex structure and learnable SE block. This overcomes the first disadvantage.

The process of extracting features of input stages by stages is not only acquisition of global context information, but also get features of organs from small to large size. Therefore, we design pyramidal structure to make full use of the above property. As the dotted box of pyramid structure in figure 1 shows, there is no information of higher level provided for the deepest part of the most abundant semantic information. Thus, we add a SE block in this position to extract features of big organs with channel attention mechanism and send them to SEC module of the shallower stage to fuse with information from the corresponding encoder. Next in this stage, we send the output of this SEC module to the decoder and shallower SEC module, which is to provide details for upsampling and syncretize features of big organs from the deeper stage and smaller organs from shallower stage, respectively. Like this, multi-size and global context information are transmitted between neighboring layers. Thus, the information flow run through the whole network from bottom to top, which overcomes the second and the third disadvantage. It provides abundant contribution for images recovering and results in more accurate segmentation as well.

D. Cascaded Auto-context Network

There are eight organs of different shapes and sizes, which makes our segmentation of OAR of NPC difficult. If we use a pre-trained network to capture the ROI, we are able to focus on the areas of targets to be segmented. Then the output is sent to another deep network, resulting in reduction of learning difficulty of this secondary network. Thus, concatenating a pre-trained network and the other one makes it easier to segment targets in ROI. Considering of this, we decide to design a cascaded network. As shown in figure 1, we use a method called auto-context for connecting two networks. The left dotted part of the SECP-Net in figure 1 is the primary network which will be pre-trained. The right dotted part of the SECP-Net in figure 1 is the secondary network. Auto-context [11] is also used for semantic segmentation. It combines original input image data and the output probability distribution and then sends the fusion into the network. With this iteration, the final segmentation results will be more accurate.

E. Comparison

Attention U-Net introduces the attention gate (AG) [6] in skip-connection. In the AG, input features are scaled with the computed attention coefficients and spatial regions are selected by analyzing both the activations and contextual information provided by the gating signal [6]. It obviously improves the skip connection by attention mechanism and can highlight salient features useful for a specific task. However, Attention U-Net doesn't deal with global context information or multi-size information. Compared to Attention U-Net, our method not only uses the proposed SEC module coordinated with pyramid structure to extract global multi-size information flow, but also utilizes the fusion structure and channel-wise attention mechanism in the SEC module to improve skip-connection.

CE-Net [7] embeds the DAC block and the RMP block in the deepest part of the network to overcome the disadvantage of lacking for multi-size features extraction. The DAC block has four cascade branches with the gradual increment of the number of artous convolution, from 1 to 1, 3, and 5, then the receptive field of each branch will be 3, 7, 9, 19. Therefore, the network can extract features from different scales. The proposed RMP block gather context information with four different-size pooling kernels to overcome the disadvantage of various sizes of objects in medical image [7]. But CE-Net doesn't pay attention to the global context information. Compared to CE-Net, we consider both global context information and multi-size information in the proposed SECP-Net by adding the SEC module and designing pyramidal network. CE-Net doesn't improve the excessively simple skip-connection while we set the SEC module in skip-connection for more non-linearity and learnability.

U-Net++ fulfills the blank part of original U-Net using dense connection from low to high level stages in the network [8]. Receptive fields vary in different stages, which have various sensitivity to diverse targets. In that case, U-Net++ is able to capture features of different levels and overlie them in channel dimension. Besides, global context information from deeper stages of the network can be transmitted to shallower stages as well. By iteratively concatenation, this makes full use of global context information. At the same time, this dense structure also makes skip-connection more complex. Compared to U-Net++, our SECP-Net designs similar pyramid architecture for purpose of global multi-size information flow. Besides, the proposed SEC module improves skip-connection by channel-wise attention mechanism of SE block, which makes skip-connection more learnable. With the help of attention mechanism, SECP-Net is capable of emphasizing more contributing features in global multi-scale information flow and ignoring the useless while U-Net++ just directly gathers global context information and multi-size features without any extra process.

In our segmentation task of OAR of NPC, there are eight kinds of organs of multi shapes and sizes to be segmented. CPF-Net proposes the GPG module, in which the global information flow is transmitted to the decoder by fusing the global context information form higher stages. The GPG module reconstruct skip-connection as well to make it more complex. In CPF-Net, multi-scale information is captured by SAPF module which consists of three parallel dilated convolution layers and dynamically fused by two scale-aware modules [9]. CPF-Net has rough extraction of multi organs' feature information due to inadequate dilated convolution kernels of various dilation rate. The proposed SECP-Net directly utilizes and fuses the multi-size information from different stage of the network, which can avoid the above disadvantage.

In summary, Attention U-Net [6], CE-Net [7] and U-Net++ [8] all aren't able to completely overcome the disadvantages of U-Net. Although CPF-Net [9] is able to improve all the three shortcomings of U-Net, CPF-Net has a disadvantage of rough extraction in our segmentation task. SECP-Net not only overcomes all the three disadvantages of U-Net, but also meets the requirement of segmentation for OAR of NPC. Furthermore, compared to the all above methods, our auto-context concatenation improves the segmentation performance by introducing the probability distribution to fix position of ROI.

F. Implementation Details

In our experiment, we use multi-class cross-entropy loss as loss function. To get best performance of our model, we choose to utilize strategy of learning rate decay which is described as:

$$lr = \frac{1}{1 + dr * epoch_{num}} lr_0 \#(1)$$

The $lr_0$ represents initial learning rate which is 0.01 and the epoch_num represents number of training epochs which is 100. We choose SGD as the optimizer and the bacthsize is set to 16. The implementation of proposed SECP-Net is based on platform Pytorch and Nvidia GeForce RTX2080ti GPU with 12GB memory. We will release our codes on Github.

# EXPERIMENTS AND RESULTS

A. Dataset and Preprocessing

The dataset used in this paper is about CT images of head and neck, which comes from SYSUCC. It consists of over 40,000 CT images from 356 patients and the resolution of a single image file is 512 × 512. We have made masks containing 13 kinds of labels for required segmentation targets, which include left eye, right eye, left temporal lobe, right temporal lobe, left mandible, right mandible, brain stem, parotid left, parotid right, spinal cord, submandibular left, submandibular right and thyroid gland. We randomly divide the dataset into a training set and a test set, CT images from 285 patients are the training set, and CT images from 71 patients are the test set. We apply five-folds cross validation to the experiment, and images are resized into 256× 256 while keeping the average aspect ratio due to insufficient memory of GPU.

B. Evaluation Metrics

We choose Dice coefficient (Dice) and Jaccard coefficient (Jac) for evaluating the performance of models, which are official evaluation standard for medical image segmentation. The higher Dice and Jac are, the more similar the results of segmentation network and ground truth will be, and thus the better performance we will get.

C. Training Scheme

To effectively train cascaded network SECP-Net, we trained our network according to the common method used for cascade network. We firstly trained the primary part of the network and then the secondary part. For the primary part, we have two steps. In step 1, we trained the backbone U-Net to convergence; In step 2, we added SEC module and pyramid structure, and then optimized the model with the initialized weights from step 1. Then we fixed the parameters in the primary network and began to train the secondary part thus the whole network was fine-tuned.

D. Results

We compare the proposed network in this paper with other remarkable models based on U-shape network including U-Net [3], Attention U-Net [6], U-Net++ [8], CE-Net [7] and CPF-Net [9]. Additionally, we also perform ablation study to verify the validity of SECP module and auto-context. In contrast and ablation experiments, we take U-Net as baseline

1) Contrast Experiments

As is shown in Table 1 and Table 2, the value with black bold is the best performance corresponding to organs to be segmented. It shows that by improving certain shortcomings of U-Net, networks like Attention U-Net have better performance of every target organ than U-Net. It's obvious that compared to the others, U-Net++ and proposed SECP-Net perform better. Although for segmentation of temporal lobe, mandible and parotid left, U-Net++ performs better than SECP-Net, our novel network has achieved more excellent results for the left eight organs. Compared to the baseline, SECP-Net achieves great improvement for left eye and thyroid, which reaches 1.27% and 2.24% for Dice respectively. It also reaches 1.44% and 2.85% for Jac respectively. Besides, our method has some improvement in average (Ave) Dice and Jac as well. On the whole, in this experiment SECP-Net has the best performance.

TABLE I THE RESULTS OF CONTRAST EXPERIMENTS ON OAR OF NASOPHARYNGEAL CARCINOMA TASK FOR DICE (%, MEAN ± STANDARD DEVIATION)

| Methods | U-Net [3] | Attention | CE-Net [7] | U-Net++ [8] | CPF-Net [9] | SECP-Net |
| --- | --- | --- | --- | --- | --- | --- |

| Methods Organs | U-Net [3] | Attention U-Net [6] | CE-Net [7] | U-Net++ [8] | CPF-Net [9] | SECP-Net |
|---|---|---|---|---|---|---|
| Temporal Lobe_L | 86.55±0.51 | 88.49±0.60 | 86.01±1.44 | **89.04±0.47** | 88.17±0.75 | 88.56±0.66 |
| Temporal Lobe_R | 86.16±0.61 | 86.66±0.59 | 85.45±1.20 | **87.75±0.50** | 87.18±0.81 | 87.55±0.61 |
| Eye_L | 75.73±0.73 | 77.46±0.66 | 73.44±1.11 | 79.92±0.52 | 77.97±0.63 | **81.19±0.77** |
| Eye_R | 75.68±0.82 | 80.33±0.67 | 75.61±1.08 | 80.03±0.56 | 79.39±0.67 | **80.81±0.71** |
| Mandible_L | 86.17±0.65 | 88.08±0.53 | 84.82±0.85 | **89.38±0.57** | 87.70±0.77 | 88.27±0.58 |
| Mandible_R | 86.52±0.67 | 87.10±0.61 | 85.42±0.79 | **88.66±0.49** | 88.04±0.73 | 88.60±0.52 |
| Brainstem | 82.39±0.68 | 84.08±0.59 | 81.22±0.73 | 84.38±0.54 | 82.22±0.56 | **85.55±0.41** |
| Parotid_L | 78.40±0.78 | 79.48±0.44 | 76.17±0.77 | **80.87±0.61** | 79.61±0.48 | 80.35±0.53 |
| Parotid_R | 77.34±0.74 | 78.89±0.46 | 77.87±0.84 | 80.53±0.70 | 78.77±0.56 | **80.61±0.48** |
| Spinal cord | 88.06±0.35 | 87.94±0.41 | 86.41±0.60 | 88.41±0.38 | 87.19±0.53 | **89.77±0.29** |
| Submandibular_L | 72.32±1.13 | 74.81±0.65 | 69.46±1.21 | 75.66±0.87 | 72.81±1.07 | **76.38±0.89** |
| Submandibular_R | 72.71±1.24 | 78.13±0.67 | 70.04±1.31 | 77.64±0.89 | 73.83±1.16 | **78.19±0.85** |
| Thyroid | 69.77±0.64 | 71.99±0.71 | 68.88±0.84 | 72.57±0.59 | 70.98±0.53 | **74.81±0.39** |
| Ave | 79.83±0.73 | 81.80±0.58 | 78.52±0.98 | 82.68±0.59 | 81.19±0.63 | **83.13±0.59** |

TABLE II THE RESULTS OF CONTRAST EXPERIMENTS ON OAR OF NASOPHARYNGEAL
CARCINOMA TASK FOR JAC (%, MEAN ± STANDARD DEVIATION)

| Methods Organs | U-Net [3] | Attention U-Net [6] | CE-Net [7] | U-Net++ [8] | CPF-Net [9] | SECP-Net |
|---|---|---|---|---|---|---|
| Temporal Lobe_L | 80.20±0.46 | 81.72±0.53 | 79.56±1.23 | **82.50±0.41** | 81.43±0.67 | 82.06±0.51 |
| Temporal Lobe_R | 80.12±0.47 | 80.26±0.54 | 79.06±1.08 | **81.51±0.39** | 81.01±0.72 | 81.40±0.49 |
| Eye_L | 69.79±0.67 | 70.98±0.51 | 67.29±0.89 | 73.47±0.42 | 71.16±0.54 | **74.91±0.59** |
| Eye_R | 69.66±0.61 | 73.49±0.53 | 68.79±0.88 | 73.47±0.39 | 72.43±0.58 | **74.52±0.57** |
| Mandible_L | 80.62±0.49 | 82.24±0.42 | 78.42±0.73 | **83.80±0.48** | 81.87±0.61 | 82.74±0.51 |
| Mandible_R | 80.93±0.47 | 81.43±0.46 | 79.12±0.72 | **83.13±0.37** | 82.22±0.64 | 83.18±0.48 |
| Brainstem | 76.56±0.59 | 77.62±0.48 | 74.55±0.63 | 78.15±0.44 | 75.77±0.48 | **79.71±0.29** |
| Parotid_L | 71.19±0.64 | 71.87±0.37 | 68.67±0.66 | **73.23±0.51** | 71.73±0.41 | 73.06±0.47 |
| Parotid_R | 70.06±0.58 | 71.41±0.38 | 70.29±0.72 | 73.09±0.62 | 70.79±0.46 | **73.26±0.33** |
| Spinal cord | 80.74±0.29 | 80.10±0.29 | 78.17±0.51 | 80.92±0.28 | 78.91±0.47 | **82.91±0.19** |
| Submandibular_L | 63.72±0.89 | 65.32±0.57 | 60.06±1.13 | 66.65±0.72 | 62.99±0.98 | **67.90±0.77** |
| Submandibular_R | 64.29±0.78 | 68.63±0.59 | 60.41±1.20 | 68.83±0.75 | 63.83±1.03 | **69.37±0.74** |
| Thyroid | 60.83±0.48 | 62.33±0.63 | 58.89±0.75 | 62.89±0.46 | 60.29±0.43 | **65.74±0.28** |
| Ave | 72.98±0.57 | 74.42±0.48 | 71.02±0.86 | 75.51±0.48 | 73.42±0.62 | **76.21±0.48** |

2) Ablation Study

The results of ablation experiments can be seen in Table 3 and Table 4. Baseline represents original U-Net.

a) Baseline+concat: This represents two original U-Net cascaded as shown in Figure 3. The left U-Net is used for rough segmentation of ROI. The output of the primary U-Net is sent to the secondary network for refining final results. We can see in tables that compared to the baseline, this method has got some improvement.

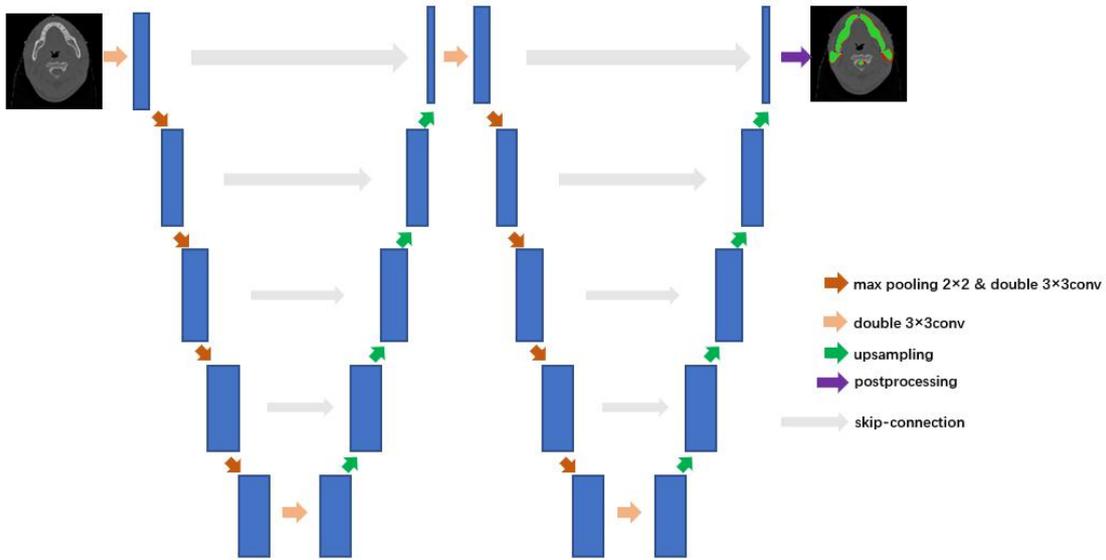

Fig.3. The structure of Baseline+concat

b) Baseline+auto-concat: This method, which is shown in Figure 4, introduces auto-context in concatenation instead of directly concatenating compared to a) Baseline+concat. The essence of image segmentation is pixel-wise classification. We combine the classification probability from the primary network and original input images. Then the combination is transmitted to the secondary network for more accurate segmentation, which achieves a better performance than direct connection.

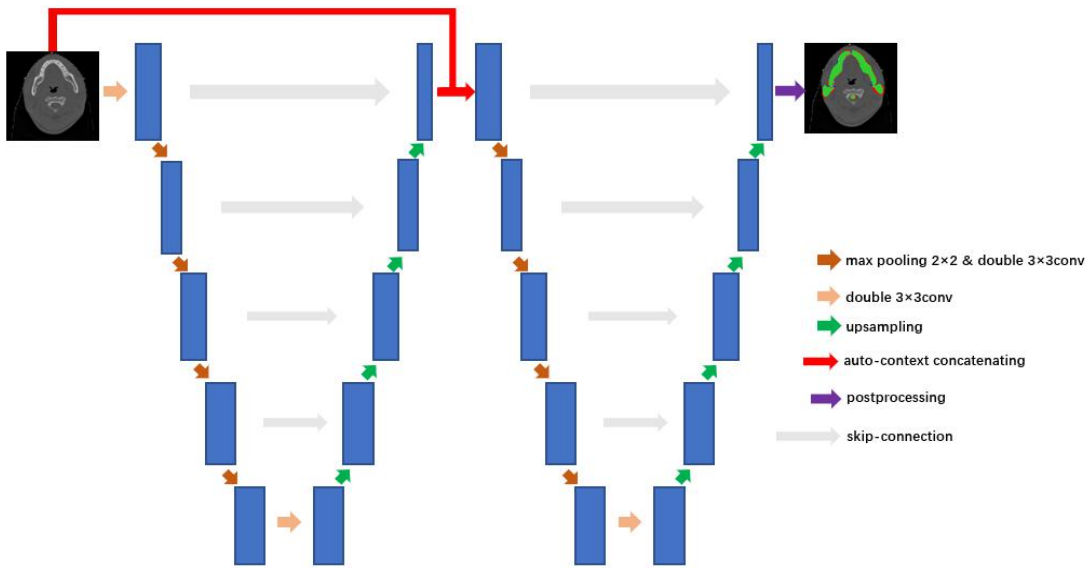

Fig.4. The structure of Baseline+auto-concat

c) Baseline+SEC: We embed SEC modules in an original U-Net as shown in Figure 5. It's obvious that global context information and multi-size information are fused successfully. Besides, more important features are selected by channel attention so that the accuracy of targets has been improved. Especially for small organs such as spinal cord, left submandibular and right submandibular, this method performs far better than baseline, which reaches 1.53%, 3.78%, 5.6% for Dice respectively and reaches 1.93%, 3.81%, 5.15% for Jac, respectively. It shows that we have great success in improving segmentation performance of small organs by capturing and fusing multi-size information.

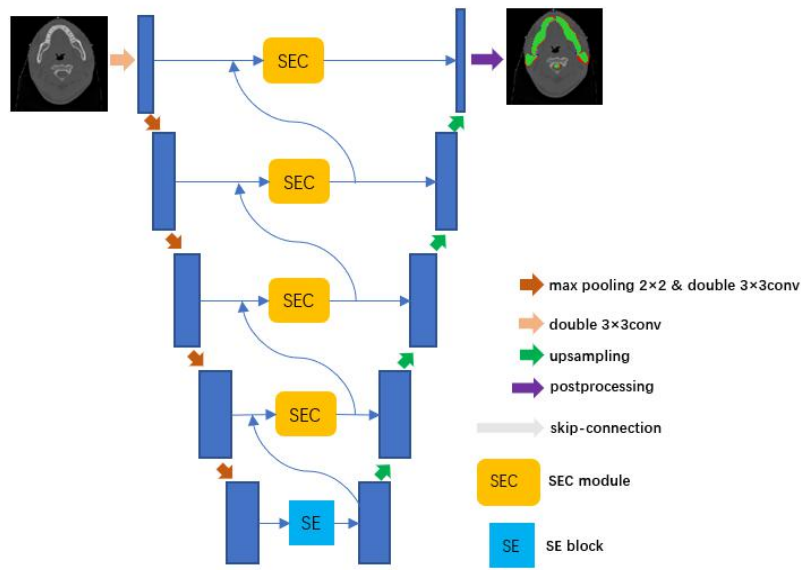

Fig.5. The structure of Baseline+SEC

d) Baseline+SEC-concat: Based on c) Baseline+SEC, we concatenate Baseline+SEC and an original U-Net as shown in Figure 6. From the results in Table 3 and Table 4, we achieve a better performance than Baseline+SEC by concatenating network and extracting syncretic global multi-size information flow. That means the combination of two methods has positive effect on segmentation.

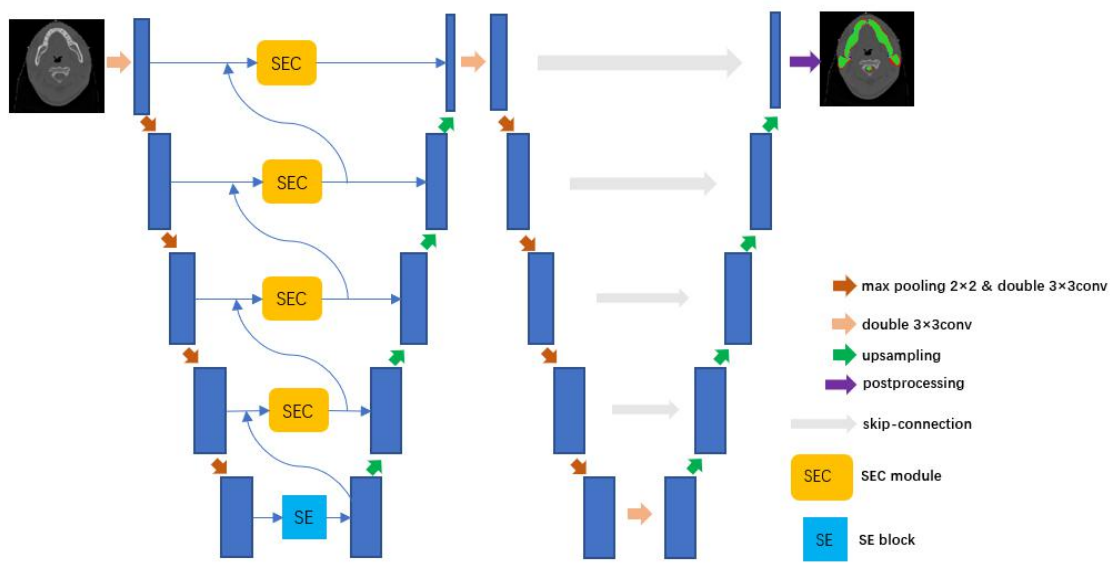

Fig.6. The structure of Baseline+SEC-concat

Finally, for results of SECP-Net on the far-right column, the proposed method in this paper introduces auto-context for concatenated network while aiming at disadvantages for U-Net like lack of multi-size information and global context information. SECP-Net achieves the best performance in our ablation experiments.

TABLE III THE RESULTS OF ABLATION EXPERIMENTS ON OAR OF NASOPHARYNGEAL CARCINOMA TASK FOR DICE (%, MEAN ± STANDARD DEVIATION)

| Methods / Organs | Bseline | Baseline+concat | Baseline+auto-concat | Baseline+SEC | Baseline+SEC-concat | SECP-Net |
|---|---|---|---|---|---|---|
| Temporal Lobe_L | 86.55±0.51 | 88.49±0.66 | 88.54±0.47 | 88.19±0.41 | 88.56±0.58 | **88.56±0.66** |
| Temporal Lobe_R | 86.16±0.61 | 87.54±0.64 | 87.52±0.53 | **87.78±0.46** | 87.78±0.60 | 87.55±0.61 |
| Eye_L | 75.73±0.73 | 79.78±0.71 | 79.49±0.74 | 79.52±0.48 | 81.06±0.53 | **81.19±0.77** |
| Eye_R | 75.68±0.82 | 79.05±0.73 | 79.21±0.69 | **81.63±0.47** | 80.30±0.49 | 80.81±0.71 |
| Mandible_L | 86.17±0.65 | 87.47±0.42 | 87.54±0.69 | 88.03±0.73 | 87.76±0.66 | **88.27±0.58** |
| Mandible_R | 86.52±0.67 | 87.33±0.56 | 88.10±0.74 | 88.55±0.76 | 88.16±0.68 | **88.60±0.52** |
| Brainstem | 82.39±0.68 | 84.06±0.38 | 85.48±0.70 | 84.72±0.78 | 85.18±0.59 | **85.55±0.41** |
| Parotid_L | 78.40±0.78 | 80.33±0.47 | 80.47±0.67 | 80.04±0.72 | 80.10±0.54 | **80.35±0.53** |
| Parotid_R | 77.34±0.74 | 79.16±0.48 | 79.45±0.61 | 80.31±0.59 | 80.20±0.52 | **80.61±0.48** |
| Spinal cord | 88.06±0.35 | 88.42±0.59 | 88.71±0.51 | 89.59±0.57 | 89.67±0.31 | **89.77±0.29** |
| Submandibular_L | 72.32±1.13 | 73.91±0.67 | 75.44±0.62 | 76.10±0.78 | 76.22±0.91 | **76.38±0.89** |
| Submandibular_R | 72.71±1.24 | 74.71±0.69 | 75.96±0.53 | **78.31±0.80** | 78.21±0.83 | 78.19±0.85 |
| Thyroid | 69.77±0.64 | 72.44±0.49 | 71.84±0.52 | 74.48±0.60 | 74.53±0.43 | **74.81±0.39** |
| Ave | 79.83±0.73 | 81.75±0.58 | 82.13±0.62 | 82.87±0.63 | 82.90±0.59 | **83.13±0.59** |

TABLE IV THE RESULTS OF ABLATION EXPERIMENTS ON OAR OF NASOPHARYNGEAL CARCINOMA TASK FOR JAC (%, MEAN ± STANDARD DEVIATION)

| Methods / Organs | Bseline | Baseline+concat | Baseline+auto-concat | Baseline+SEC | Baseline+SEC-concat | SECP-Net |
|---|---|---|---|---|---|---|
| Temporal Lobe_L | 80.20±0.46 | 82.01±0.53 | 82.04±0.39 | 81.62±0.32 | 82.03±0.47 | **82.06±0.51** |
| Temporal Lobe_R | 80.12±0.47 | 81.41±0.54 | 81.59±0.48 | **81.60±0.39** | 81.60±0.52 | 81.40±0.49 |
| Eye_L | 69.79±0.67 | 73.52±0.59 | 73.30±0.66 | 73.26±0.42 | 74.88±0.44 | **74.91±0.59** |
| Eye_R | 69.66±0.61 | 72.87±0.62 | 72.97±0.60 | **75.15±0.39** | 73.81±0.38 | 74.52±0.57 |
| Mandible_L | 80.62±0.49 | 81.84±0.33 | 81.93±0.61 | 82.46±0.58 | 82.20±0.55 | **82.74±0.51** |
| Mandible_R | 80.93±0.47 | 81.68±0.46 | 82.42±0.68 | 83.11±0.57 | 82.69±0.60 | **83.18±0.48** |
| Brainstem | 76.56±0.59 | 78.05±0.33 | 79.52±0.63 | 78.87±0.65 | 79.33±0.48 | **79.71±0.29** |
| Parotid_L | 71.19±0.64 | 72.99±0.42 | 73.06±0.56 | 72.77±0.61 | 72.81±0.41 | **73.06±0.47** |
| Parotid_R | 70.06±0.58 | 71.72±0.40 | 72.03±0.49 | 72.99±0.52 | 72.85±0.46 | **73.26±0.33** |
| Spinal cord | 80.74±0.29 | 81.09±0.48 | 81.38±0.44 | 82.67±0.46 | 82.82±0.24 | **82.91±0.19** |
| Submandibular_L | 63.72±0.89 | 65.13±0.57 | 66.55±0.48 | 67.53±0.69 | 67.74±0.78 | **67.90±0.77** |
| Submandibular_R | 64.29±0.78 | 66.11±0.59 | 67.24±0.45 | **69.44±0.72** | 69.42±0.72 | 69.37±0.74 |
| Thyroid | 60.83±0.48 | 63.21±0.43 | 62.68±0.41 | 65.17±0.46 | 65.20±0.32 | **65.74±0.28** |
| Ave | 72.98±0.57 | 74.74±0.48 | 75.13±0.53 | 75.90±0.52 | 75.95±0.49 | **76.21±0.48** |

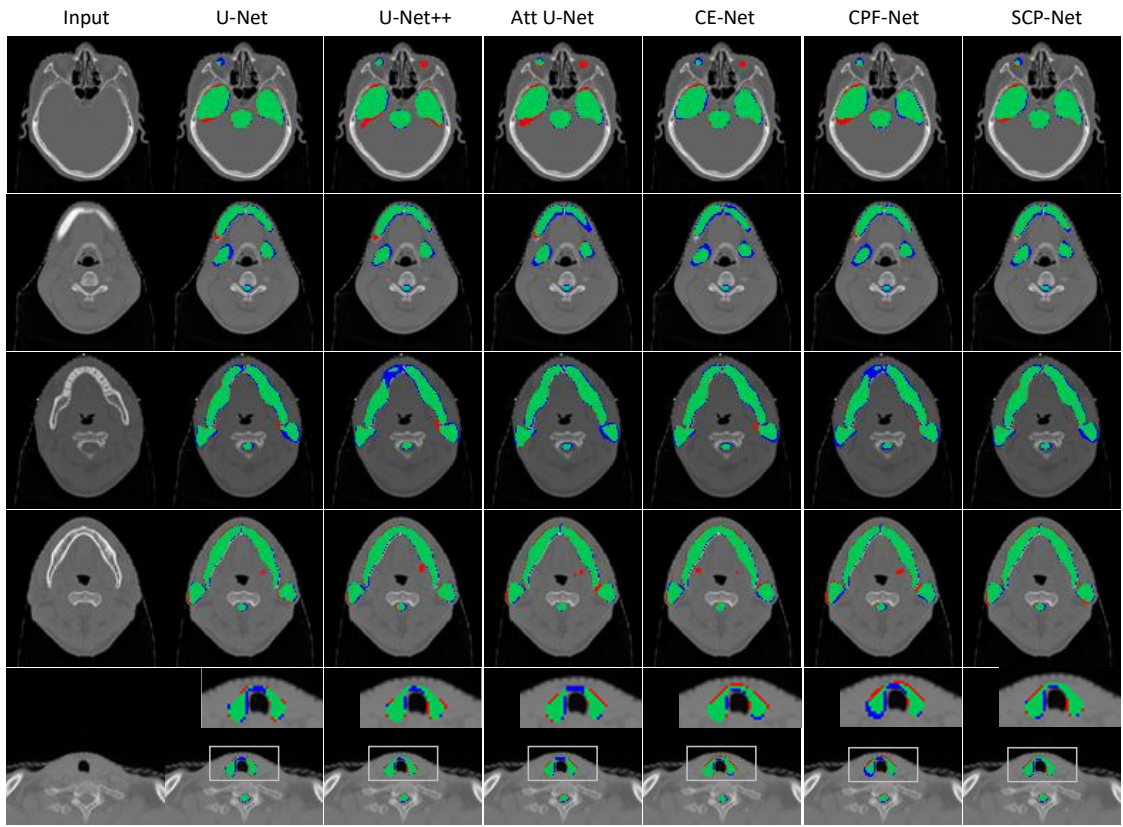

Fig. 7 Visualization results of segmentation experiments. The blue part represents ground truth, red part is the results predicted by each model and green part is intersection of ground truth and the segmented results by model (The Ture Positive). From left to right: original image, U-Net, U-Net++, Attention U-Net, CE-Net, CPF-Net and our SECP-Net.

Figure 7 shows the visualization results of different models. SECP-Net also provides the best segmentation results based on this subjective index.

In the first row of the figure, eyes, temporal lobe and brain stem are segmented from up to down in images by above models. We can easily see that U-Net++, Attention U-Net, CE-Net and CPF-Net all give wrong segment results for left eye because actually left eye doesn't exist in this single slice. Neither of U-Net and our SECP-Net make such a mistake while our SECP-Net has more accurate segmentation on right eye and temporal lobe.

In the second row of the figure, mandible, submandibular, spinal cord are segmented from top to bot. Attention U-Net and CE-Net both have terrible segmentation of mandible distinctly while U-Net++, Attention U-Net and CPF-Net all give redundant segmentation part of mandible (the red part). Our SECP-Net gives a more accurate result of both mandible and submandibular due to more intersection part.

In the third row of the figure, mandible, parotid and spinal cord are segmented from the top down. It's apparent that both U-Net++ and CPF-Net have poor segmentation results of mandible while the other methods don't. Compared to U-Net, Attention U-Net and CE-Net, our SECP-Net has got more accurate segmentation for left parotid due to more overlap part of ground truth and the predicted.

In the fourth row of the figure, mandible, parotid and spinal cord are segmented from the top down. In fact, submandibular doesn't exist in this image. However, nearly all methods give wrong prediction about submandibular (the red part in the middle position of images) except the proposed SECP-Net. It's obvious that SECP-Net performs far better than above models when it comes to small organs named submandibular. It's effective to capture multi-size information by pyramid structure, select useful features by channel attention mechanism and send syncretic information flow through skip connection to upsample for improving accuracy of segmentation.

In the last row of the figure, thyroid and spinal cord are segmented from up to down. Apart from CE-Net and SECP-Net, the other methods give disconnected results instead of actually contiguous ones of thyroid (the blue part between left and right part of thyroid). SECP-Net has more exact segmentation than CE-Net because of more intersection part.

## CONCLUSION

Since traditional U-shape networks have an intuitive ship-connection, its performance is easily distorted by some noise due to the weakness of learning ability. Moreover, U-shape networks cannot extract the multi-size information, and lacks use of the global context information. For overcoming these disadvantages of U-shape networks, a novel pyramidal deep learning model named SECP-Net is proposed to automatically segment the OAR of NPC in CT images. In SECP-Net, SE-Connection module and pyramid structure is used to capture global multi-size information flow. The channel attention mechanism is fully utilized to highlight contributing features, and the global context information is applied when concatenating networks with auto-context.

A CT images dataset, which is provided by SYSUCC, is used to evaluate the performance of SECP-Net. Comparing with other competitive models, i.e., U-Net [3], Attention U-Net [6], U-Net++ [8], CE-Net [7] and CPF-Net [9], the experimental results show that the proposed SECP-Net can outperformance than other competitive models in this dataset. Moreover, the designed SEC module, pyramid structure and auto-context concatenation can be proved as successful and effective parts for the OAR segmentation during ablation study. Except for the channel attention, the local region attention is needed to further research, especially establishing the relationship between the foreground and background. In future, the image segmentation method based on U-shape networks will be combined with the transformer architecture to further improve the system performance.